\documentclass[aps,prc,preprint,amsmath,showpacs,superscriptaddress]{revtex4}
\usepackage{graphicx,epsfig}
\newcommand{\beqy}{\begin{eqnarray}}
\newcommand{\eeqy}{\end{eqnarray}}
\newcommand{\bmlet}{\begin{subequations}}
\newcommand{\emlet}{\end{subequations}}
\newcommand{\bfdel}{\mbox{\boldmath$\nabla$}}
\begin{document}
\title{Semi-classical equation of state and specific heats for neutron-star 
inner crust with proton shell corrections.}
\author{M. Onsi}
 \affiliation{D\'ept.\ de Physique, Universit\'e de Montr\'eal,
Montr\'eal (Qu\'ebec), H3C 3J7 Canada}
\author{A. K. Dutta}
\affiliation{D\'ept.\ de Physique, Universit\'e de Montr\'eal, 
Montr\'eal (Qu\'ebec), H3C 3J7 Canada}
\affiliation{School of Physics, Devi Ahilya University, Indore 452001, India }
\author{H. Chatri}
 \affiliation{D\'ept.\ de Physique, Universit\'e de Montr\'eal,
Montr\'eal (Qu\'ebec), H3C 3J7 Canada}
\author{S. Goriely}
\affiliation{Institut d'Astronomie et d'Astrophysique,
Universit\'e Libre de Bruxelles - CP226, 1050 Brussels,  Belgium}
\author{N. Chamel}
\affiliation{Institut d'Astronomie et d'Astrophysique,
Universit\'e Libre de Bruxelles - CP226, 1050 Brussels,  Belgium}
\author{J. M. Pearson}
\affiliation{D\'ept.\ de Physique, Universit\'e de Montr\'eal,
Montr\'eal (Qu\'ebec), H3C 3J7 Canada}
\begin{abstract}

An approach to the equation of state for the inner crust of neutron stars based on Skyrme-type forces is presented. Working within the Wigner-Seitz picture,
the energy is calculated by the TETF (temperature-dependent extended 
Thomas-Fermi) method, with proton shell corrections added self-consistently
by the Strutinsky-integral method.
Using a Skyrme force that has been fitted to both neutron matter and to
essentially all the nuclear mass data, we find strong proton shell effects: 
proton numbers $Z$ = 50, 40 and 20 are the only values possible in the inner 
crust, assuming that nuclear equilibrium is maintained in the cooling neutron
star right down to the ambient temperature. 

Convergence problems with the TETF expansion for the entropy, and our way of 
handling them, are discussed. Full TETF expressions for the specific heat of
inhomogeneous nuclear matter are presented. Our treatment of the electron gas, 
including its specific heat, is essentially exact, and is described in detail.
\end{abstract}


\maketitle\thispagestyle{empty}
\newpage\setcounter{page}{2}
\section{Introduction}
\renewcommand{\theequation}{1.\arabic{equation}}
\setcounter{equation}{0}
\label{intro}
  
We are concerned with the application of Skyrme-type effective nuclear 
interactions to the determination of the equation of state (EOS) of the 
inhomogeneous nuclear matter encountered at nuclear and subnuclear densities 
in core-collapse supernovas and in the inner crust of neutron stars. In our 
first paper on this topic \cite{opp97} we adopted a Wigner-Seitz (WS) model of 
the inhomogeneous nuclear medium, and used the 
fourth-order semi-classical temperature-dependent extended Thomas-Fermi (TETF) 
method to calculate the kinetic energy and entropy. That paper dealt primarily 
with the conditions prevailing in core-collapse supernovas. The present
paper relates rather to the inner crust of neutron stars, describing in
particular some modifications to the earlier model, made necessary by two
problems that emerge at the much lower temperatures $T$ that are involved: 
i) the TETF expansion (in powers of $\hbar^2$) for the entropy converges badly 
at low $T$; ii) proton shell effects are not negligible at low $T$. 

This last point is especially important if one is interested in the 
neutron-star crust as a possible alternative site for the synthesis of the 
so-called r-process elements \cite{latt77,mey89,frei99,gor05}. The usual model 
of the r-process of nucleosynthesis is associated with the {\it birth} of a 
neutron star in a core-collapse supernova, during which ``seed" nuclei are 
exposed to an intense flux of neutrons. Rapid (``$r$") capture of neutrons 
alternating with beta decay leads to the formation of a string of highly 
neutron-rich isotopes of a wide range of elements, which, once the source of 
neutrons is removed, will beta-decay back to the most neutron-rich stable 
isobar for the given mass number $A$ (see Ref. \cite{agt07} for a recent 
review). The alternative picture, of interest here, is associated rather with 
the {\it death} of a neutron star, or at least with its partial disruption.
Because of the very large densities, the matter in a neutron star is highly 
neutron rich, and the closer to the center
the more neutron-rich it will be. But if for one reason or another matter
is ejected from the neutron star it will rapidly decompress, and so will be
able to undergo a chain of beta decays, the end-product of which will again be
r-process nuclei. Ejection of matter from a neutron star is usually supposed
to result from the merger of one neutron star with another, or with a black
hole \cite{frei99,rj01}, but other scenarios have been envisaged, e.g.,
volcanoes \cite{dys69}, magnetars \cite{hl06}, quark stars \cite{jai07} and
explosions resulting from the mass of the neutron star falling below the 
minimal critical value \cite{sum98}. However, the precise ejection
mechanism is of no concern to us in this paper.

It is convenient at this point to
recall that at least three distinct regions can be recognized in a neutron 
star: a central, locally homogeneous, core, and two concentric
shells characterized by different inhomogeneous phases \cite{pr95}. The
outermost of these shells, the ``outer crust", consists of an electrically
neutral lattice of nuclei and electrons. At the surface of the star only nuclei
that are stable under natural terrestrial conditions are found (in fact, 
nuclear equilbrium, discussed below, implies that  only $^{56}$Fe will be 
found), but on
moving towards the interior the increasing density leads to the appearance
of nuclei that are more and more neutron rich, until at a mean local density
$\bar{\rho}$ of around 2.4 $\times 10^{-4}$ nucleons.fm$^{-3}$
(4.0 $\times 10^{11}$ g.cm$^{-3})$ neutron drip sets in. This marks the
transition to the ``inner crust", which at least up to a mean density of
$\bar{\rho}$ = 0.06 nucleons.fm$^{-3}$ consists of neutron-proton clusters,
or droplets, immersed in a neutron gas, with the neutralizing electron gas
being essentially uniform (we neglect screening effects in this paper).
It is equally well established that by the point where the mean density has
risen to around $\bar{\rho}$ = 0.10 nucleons.fm$^{-3}$, i.e., about 2/3 of the
density $\rho_0$ of symmetric infinite nuclear matter (INM) at equilibrium,  
the droplet phase no longer
exists and has been replaced by the homogeneous phase of the core, which
consists primarily of neutrons, with a small admixture of proton-electron
pairs, and possibly other particles, including free quarks, closer to the
center.

What happens in the transition region over the range
0.06 $\le \bar{\rho} \le $ 0.10 nucleons.fm$^{-3}$, close to the inside edge 
of the inner crust, is far less clear. The question cannot be settled by
observation at the present time, and theoretical predictions are sensitive to
the details of the calculations, in particular to the choice of the effective 
interaction. For some interactions the transition from the droplet 
phase to the homogeneous phase is indirect and complex, with a whole sequence 
of different inhomogeneous phases being formed. At the interface with the 
homogeneous core these calculations find a ``bubble" phase, this taking the 
form of bubbles of neutron gas in a denser liquid of neutrons and protons, the 
droplet phase having effectively been turned inside out. Furthermore, at 
slightly lower densities, between the bubble and droplet phases, several 
so-called ``pasta" phases are predicted to put in an appearance, these
being characterized by exotic, non-spherical shapes \cite{pr95}. On the other
hand, it has been shown that for other effective interactions the situation is
much simpler, with no bubble or pasta phases being formed (at least at the
assumed zero temperature of a stable neutron star): at a mean density of
around $\bar{\rho}$ = 0.075 nucleons.fm$^{-3}$ the droplet phase undergoes a
transition directly to the homogeneous phase (see Ref. \cite{dh00}, and
references cited therein; also Ref. \cite{mar05}). In the present paper 
paper we will avoid these ambiguities by limiting ourselves to values of
$\bar{\rho}$ less than 0.06 nucleons.fm$^{-3}$, which means that we would not
be able to deal with ejection mechanisms that reached even deeper into the 
star.

Since neutron stars are formed at temperatures of the order of 10 MeV 
(10$^{11}$ K) and rapidly cool to around 0.1 MeV \cite{pr95}, it is usually 
assumed that the final composition of the stable star corresponds to nuclear 
and beta equilibrium at a temperature of $T$ = 0, the configuration of 
so-called ``cold catalyzed matter"; we shall later examine the validity of this
assumption. Determining the composition of the outer crust in 
this picture is straightforward (see, for example, Ref. 
\cite{rhs06}): the equilibrating nucleus at each given density (or pressure) is
found from the known nuclear masses, as given by experiment or, where mass 
data are unavailable, a mass model such as the FRDM \cite{frdm} or 
HFB-14 \cite{gsp07} (see also Refs. \cite{lpt03,pg06} for reviews). We shall
therefore not consider the outer crust any further here.

As for the composition of the inner crust of the stable neutron star,
the relevant question at a given mean density $\bar{\rho}$ is to determine the
total number of neutrons $N$, including those of the vapor, and protons $Z$
per cluster. For this one needs the total Helmholtz free energy per nucleon 
$f$ (including the electronic contribution) at the ambient temperature
(usually taken to be zero), as a function of the density and the composition 
$X \equiv (Z, A = Z + N)$; one then 
minimizes $f$ with respect to $N$ and $Z$ at constant $\bar{\rho}$. 
(Alternatively, to determine the composition at a given {\it pressure} $P$ one 
minimizes the Gibbs free energy per nucleon $g$ with respect to $N$ and $Z$ 
at constant  $P$. It follows from the easily proven thermodynamical relation
\beqy\label{1.1}
\Big(\frac{\partial g}{\partial X}\Big)_{P,T} =
\Big(\frac{\partial f}{\partial X}\Big)_{\bar{\rho},T} 
\eeqy
that the two procedures are completely equivalent. We nevertheless find it more
convenient to work with the Helmholtz free energy $f$ at given values of 
$\bar{\rho}$: see, for example, Section II of Ref. \cite{bps71}.) 

The pressure in a layer of the crust of density $\bar{\rho}$ can then
be found by numerical differentiation from the identity
\beqy\label{1.2}
P = \bar{\rho}^2\Big(\frac{\partial f}{\partial \bar{\rho}}\Big)_ {T,X} \quad ;
\eeqy 
note particularly that $f$ is a mean quantity, averaged over 
inhomogeneities, and not a local quantity. With the pressure $P$ determined as 
a function of the mean 
density $\bar{\rho}$, the values of $P$ and $\bar{\rho}$ in any layer of
the neutron star, along with the local composition, can be determined 
through solution of the Tolman-Oppenheimer-Volkoff equation \cite{tol39,ov39}.

When, for one reason or another, decompression of crustal material begins,
the temperature may start to rise. To follow the evolution of this process
we shall require the EOS for non-zero values of $T$, and also the specific
heat per nucleon at constant volume, $c_v$, given in terms of the entropy
per nucleon, $s$, by 
\beqy\label{1.3}
c_v = T\Big(\frac{\partial s}{\partial T}\Big)_{\bar{\rho},X} 
\quad .
\eeqy
The entropy itself is given in terms of the Helmholtz free energy by
\beqy\label{1.4}
s = -\Big(\frac{\partial f}{\partial T}\Big)_{\bar{\rho},X} \quad .
\eeqy
Thus all quantities of interest here can be derived from a calculation of
$f$ as a function of $\bar{\rho}$ and $T$. Note that $s$ and $c_v$, like $f$, 
are mean quantities. 

A popular EOS that has been extensively applied to supernova explosions is
that of Lattimer and Swesty~\cite{ls91}. However, the applicability of
this EOS to neutron-star crusts is limited by the fact that it is based on
the so-called compressible liquid-drop model without any shell corrections,
which at the prevailing low temperatures can be expected to be significant.
Actually, both Refs. \cite{latt77} and \cite{mey89} attempt to take account of
shell effects, although in a rather rudimentary way, by making use of the
algebraic bunching technique of Myers and Swiatecki \cite{ms66}. 

In the present paper, as in Ref. \cite{opp97}, we model the inhomogeneous
nuclear medium by a single spherical WS cell, and attempt to incorporate shell
effects into this framework microscopically and self-consistently, thereby
permitting some measure of continuity of treatment across the interface between
the inner and outer crusts. The most obvious way to do this is through the
Hartree-Fock (HF) method, as has already been done, for example, by Bonche and Vautherin
\cite{bv81} at finite temperature and by Negele and Vautherin\cite{nv73} at zero temperature, 
using the WS approximation. However, we abandoned this approach
for the following reason. While protons are strongly bound in the inner crust
because of the large neutron excess, and thus show strong shell effects, for
neutrons, by the very definition of the inner crust, 
there will be a continuous spectrum of
unbound single-particle (s.p.) neutron states that are occupied. Thus any
neutron added to the system must in reality go into this continuum, whence it 
follows that we should not expect any neutron shell effects. Actually, this 
conclusion will hold only if the dripped neutrons form a uniform liquid, and in
reality scattering of unbound neutrons on the inhomogeneities of the
crust may give rise to so-called Casimir or band effects\cite{bm01,mh02,mbh03},
whose exact evaluation requires the application of the band theory of solids
(see Ref. \cite{cha07}
and references quoted therein). Nevertheless, these neutron shell
effects are much smaller than the proton ones \cite{oy94} and have therefore a
negligible impact on the EOS and the equilibrium composition of the inner
crust, although they are known to be significant for transport properties
\cite{cha06}. 
However, in practice any HF calculation in the WS approximation
involves discretization, giving rise to shell effects for both protons and
neutrons. But, as we have argued above, these neutron shell effects must be 
spurious, and in the HF calculation of Ref. \cite{nv73} special steps had to be
taken to smooth them (see also Refs. \cite{bst06,cha07}). We conclude that
as far as neutrons are concerned the semiclassical extended Thomas-Fermi method
is better adapted to a WS approach than is the HF method.

The solution we adopt here to the problem of including the appropriate proton 
shell corrections without introducing spurious neutron shell corrections is to 
use the ETFSI (extended Thomas-Fermi plus Strutinsky integral) high-speed 
approximation to the HF method \cite{dut,ton,pea,abo1,abo2}.
We have already made an exploratory study
of the applicability of this method to a WS picture of the EOS, and found that
proton shell effects are indeed important \cite{dop04}, but here, in addition
to making much more extensive calculations of the EOS, we improve the $T > 0$
results by taking account of possible shell effects in the entropy, which will
manifest themselves in the free energy through the relation 
\beqy\label{1.5}
f = e - Ts  \quad  ,
\eeqy
where $e$ is the energy per nucleon. 

A further development of considerable significance is that the TETFSI method,
as we shall refer to this temperature-dependent ETFSI method, is no longer
limited to forces whose effective nucleon mass $M^*$ is equal to the real
mass $M$. This permits us to use more realistic effective forces with smaller
values of the effective mass. Thus in the present calculations the effective
interaction that we use is the Skyrme force BSk14, for which the effective mass
in symmetric INM at the equilibrium density $\rho_0$ (0.159 nucleons.fm$^{-3}$)
is 0.800$M$, which is to be compared with the value of 0.825$M$ found in 
extended Brueckner-Hartree-Fock calculations that include three-nucleon forces
\cite{cao06}. This is the force that underlies the Hartree-Fock-Bogoliubov 
(HFB) mass model HFB-14 \cite{gsp07}, a force that is
eminently suitable for calculating the properties of neutron-star crustal
matter, since on the one hand it has been fitted to the properties of neutron
matter, as determined by calculations with realistic two- and three-nucleon
forces \cite{fp81}, and on the other hand it gives an excellent fit to
essentially all the available mass data ($\sigma_{rms} = 0.729$ MeV). Given 
that the neutron-star crust is both inhomogeneous and contains some protons, 
the high quality of the mass fit is especially relevant, since it means a) that
inhomogeneities in nuclear matter (surface effects in droplet-model language) 
are well modeled, and b) that neutron-proton interactions are well represented.
(However, no Skyrme force should be used for the highly supernuclear densities 
encountered deep within the core of a neutron star.) We stress that in this
paper we neglect pairing, as in Refs. \cite{nv73,bv81}. 

In Section II we discuss our parametrization of the WS cell. Section III 
describes our adaptation of the ETFSI method to the problem of the EOS of the
neutron-star inner crust at non-zero temperatures, with particular attention
to the convergence properties of the ETF expansion of the entropy. Our
formalism is applied in Section IV to the properties of the inner crust of a 
neutron star (we do not examine in this paper the
important question of the rapid decompression of neutron-star matter). The 
existence of strong proton-shell effects in the inner crust is discussed in 
this same section. Our conclusions are summarized in Section V. Some important
material is to be found in the appendices, notably 
the TETF expansion of the specific heat (App. A) and a proof of the 
Strutinsky-integral theorem (App. C).

\section{The Wigner-Seitz cell}
\label{wscell}
\renewcommand{\theequation}{2.\arabic{equation}}
\setcounter{equation}{0}

Since we do not consider depths greater than that for which the mean density is
$\bar{\rho}$ = 0.06 nucleons.fm$^{-3}$ only the droplet phase of nucleons has 
to be considered, and we shall assume that the WS cell associated with this 
phase is spherical, radius $R_c$. This cell is entirely representative of the
macroscopically sized volume element being considered, in the sense that all
the nucleons of this volume element are imagined to be grouped into identical 
such cells, there being one cell for each droplet. The average neutron and 
proton densities over the locally representative cell must thus each be equal 
to the local values of the corresponding macroscopic densities, $\bar{\rho}_n$ 
and $\bar{\rho}_p$, given respectively by
\bmlet
\beqy\label{2.1a}
\bar{\rho}_n = \bar{\rho}(1-Y_e)
\eeqy
and
\beqy\label{2.1b}
\bar{\rho}_p = \bar{\rho}Y_e   \quad  ,
\eeqy
\emlet
where $Y_e$ is the fraction of nucleons that are protons. The neutron and 
proton density distribution functions within the cell, $\rho_n({\bf r})$ and 
$\rho_p({\bf r})$, are then constrained by
\beqy\label{2.2}
\frac{\int_{cell} \rho_q({\bf r})\,d^3{\bf r}}{\int_{cell}\,d^3{\bf r}} =
\bar{\rho}_q = \frac{3}{R_c^3} \int_0^{R_c}\rho_q(r)r^2\,dr \quad ,
\eeqy
where $q$ denotes $n$ or $p$, as the case may be, and the second equality holds
in the case of spherical symmetry, assumed here. The total number of nucleons
of each type in the cell is
\beqy\label{2.3}
N_q = \frac{4\pi}{3}R_c^3 \,\bar{\rho}_q 
\eeqy 
($N_n = N, N_p = Z$).

For the neutron and proton density distribution functions we adopt a modified
version of the simple Fermi form that we used in Ref. \cite{opp97}: limiting
ourselves to the spherical case and writing
\beqy\label{2.4}
\rho_q(r) = \rho_{Bq} + \rho_{0q}f_q(r)  \quad ,
\eeqy
in which $\rho_{Bq}$ is the usual constant background term, we now take
\beqy\label{2.5}
f_q(r) = \frac{1}{1 + \frac{1}{e}\,\exp \Big(\frac{C_q - R_c}
{r - R_c}\Big)^2\, \exp \Big(\frac{r-C_q}{a_q}\Big) }\quad .
\eeqy
Here the denominator of the radially varying term contains an extra factor 
$\frac{1}{e}\,\exp \Big(\frac{C_q - R_c}{r - R_c}\Big)^2$, the presence of
which guarantees that the density gradient now vanishes at the surface of the 
cell, thereby ensuring that the droplets merge smoothly with the homogeneous
neutron vapor within the WS cell (in addition to being physically realistic, 
this condition is also required for 
the validity of the semi-classical part of the TETFSI method used here to 
calculate the nuclear kinetic energy and entropy). Also, again because of the 
extra factor, $f_q(r)$ itself vanishes on the surface of the WS cell, whence
\beqy\label{2.6}
\rho_q(R_c) = \rho_{Bq} \quad.
\eeqy
The factor $1/e$ ensures that the radially varying factor 
takes the value $\rho_{0q}/2$ at $r = C_q$, as usual. Finally, we note that
$f_q(r)$ varies monotonically {\it over the cell}.

The two background constants $\rho_{Bq}$ are not independent of the other 
parameters, but rather are fixed by
\beqy\label{2.7}
\rho_{Bq} = \bar{\rho}_q -\frac{3}{R_c^3}I_q\rho_{0q}  \quad  ,
\eeqy
where
\beqy\label{2.8}
I_q = \int_0^{R_c}f_q(r)r^2\,dr \quad .
\eeqy
For given $\bar{\rho}$ the WS cell is thus characterized by seven geometrical 
parameters, $\rho_{0q}$, $C_q$, $a_q$ ($q = n, p$), and $R_c$, in addition to 
the composition parameter $Y_e$. In the most general case all eight of these
parameters correspond to degrees of freedom, but either or both of the last two
might be constrained to fixed values, according to the physical situation
being described. In particular, for specified $Z$ and $N$, the given value of
$\bar{\rho}$ determines $R_c$.

But in all cases the complete set of parameters is subject to the 
additional constraint that the densities $\rho_q(r)$ must be positive at all 
points in the cell. We handle this problem as described in Section II of 
Ref. \cite{opp97}, the modified radial distribution leading to the simplified 
condition
\beqy\label{2.9}
\frac{\bar{\rho_q}}{(3I_q/R_c^3) - f_q(r=0)}  < \rho_{0q} 
< \frac{R_c^3\,\bar{\rho}_q}{3I_q}  \quad  ,
\eeqy
given that $f_q(r)$ varies monotonically over the cell.
The lower limit is seen from the mean-value theorem to be essentially negative,
and since negative values of $\rho_{0q}$ correspond to bubbles we shall be
interested here only in the restricted range
\beqy\label{2.10}
0 < \rho_{0q} < \frac{R_c^3\,\bar{\rho}_q}{3I_q}  \quad . 
\eeqy

Although this is of no concern for the present paper, it is convenient to 
note here that once decompression and beta decay have begun, it will be
necessary to constrain the background parameters $\rho_{Bq}$ to given values, 
since different ($Z, A$) configurations will be present simultaneously. Thus,
with Eq. (\ref{2.7}) still holding, the two degrees of freedom corresponding to
the parameters $\rho_{0q}$ will be lost. Rather, they will be determined
uniquely according to
\beqy\label{2.11}
\rho_{0q} = \frac{R_c^3}{3I_q} (\bar{\rho}_q - \rho_{Bq}) \quad ,
\eeqy
positive values still corresponding to droplets and negative to bubbles.
The condition that the densities be everywhere positive then leads to 
constraints on the range of values that are possible for $\rho_{Bq}$: 
\beqy\label{2.12}
0 < \rho_{Bq} < \frac{\bar{\rho}_q}{1 - 3I_q/\{R_c^3f_q(r=0)\}} \quad,
\eeqy
in which the upper limit is essentially larger than $\bar{\rho}_q$, but
only the restricted range
\beqy\label{2.13}
0 < \rho_{Bq} < \bar{\rho}_q 
\eeqy
corresponds to droplets.

\section{The TETFSI method}
\renewcommand{\theequation}{3.\arabic{equation}}
\setcounter{equation}{0}
\label{method}

For a given set of the geometrical cell parameters we first write the 
total density of the Helmholtz free energy at a given point in the cell as
\beqy\label{3.1}
{\mathcal F}^{\,\prime} = {\mathcal F}_{nuc} + {\mathcal F}_{e} + 
{\mathcal E}_c + (\bar{\rho}_nM_n + \bar{\rho}_pM_p + n_em)c^2 \quad ,
\eeqy
where ${\mathcal F}_{nuc}$ is the specifically nuclear free-energy density
(discussed below), ${\mathcal F}_{e}$ is the density of the electron free 
energy (calculated essentially exactly), and ${\mathcal E}_c$ is the Coulomb 
energy. (Strictly speaking, in
this equation $\bar{\rho}_n$ and $\bar{\rho}_p$ should be replaced by $\rho_n$ 
and $\rho_p$, respectively, but the differences vanish on integrating over the 
cell.) Since $\bar{\rho}_p = n_e$, electrical neutrality holding
globally over the WS cell, we can now write 
\beqy\label{3.2}
(\bar{\rho}_nM_n + \bar{\rho}_pM_p + n_em)c^2 = 
\bar{\rho}\Big\{(1 - Y_e)M_n + Y_e(M_p + m)\Big\}c^2=
\bar{\rho}M_nc^2 - \bar{\rho}Y_eQ_{n,\beta}  \quad ,
\eeqy
where $Q_{n,\beta}$ is the beta-decay energy of the neutron (0.782 MeV). But 
the term $\bar{\rho}M_nc^2$ makes a constant contribution to the free energy 
per nucleon and can thus be discarded. Thus in place of Eq. (\ref{3.1}) we 
write
\beqy\label{3.3}
{\mathcal F} = {\mathcal F}_{nuc} + {\mathcal F}_{e} + {\mathcal E}_c
- \bar{\rho}Y_eQ_{n,\beta} \quad .
\eeqy
The total Coulomb energy density, direct and exchange, is given in  general by
\beqy\label{3.4}
{\mathcal E}_c({\bf r}) = \frac{e^2}{2}\{\rho_p({\bf r}) - n_e\}
\int\frac{\rho_p({\bf r}^\prime) - n_e}
{|{\bf r} - {\bf r^{\prime}}|}d^3{\bf r^{\prime}}
- \frac{3e^2}{4} \Big(\frac{3}{\pi}\Big)^{1/3}(\rho_p^{4/3} + n_e^{4/3})
\quad ,
\eeqy
where for the exchange term we have used the usual Kohn-Sham variant 
\cite{ks65} of the Slater approximation. The last equation 
reduces in the case of spherical symmetry to
\beqy\label{3.5}
{\mathcal E}_c = 2\pi e^2(\rho_p - n_e)\Big\{\int_0^r \rho_p(r^\prime)
\Big(\frac{{r^\prime}^2}{r}-r^\prime\Big)\,dr^\prime +
n_e\frac{r^2}{6}\Big\}       \\ \nonumber
- \frac{3e^2}{4} \Big(\frac{3}{\pi}\Big)^{1/3}(\rho_p^{4/3} + n_e^{4/3})
\quad .
\eeqy

The mean free energy per nucleon in the entire system is given by the 
corresponding quantity averaged over just one cell,
\beqy\label{3.6}
f = \frac{1}{A}\int{\mathcal F}({\bf r})d^3{\bf r} \quad  ,
\eeqy
where $A = N + Z$ is the total number of nucleons in the cell. 
We shall likewise calculate the total entropy, the density of which at a given
point can be written as
\beqy\label{3.7}
\sigma = \sigma_{nuc} + \sigma_e  = \sigma_n + \sigma_p + \sigma_e\quad ,
\eeqy
whence for the mean total entropy per nucleon we have
\beqy\label{3.8}
s = \frac{1}{A}\int\sigma({\bf r})d^3{\bf r}    \quad .
\eeqy

The Skyrme force BSk14 \cite{gsp07} for which we calculate the
densities ${\cal F}_{nuc}$ and $\sigma_{nuc}$ has the usual form
\beqy\label{3.9}
v_{ij} & = & t_0(1+x_0P_\sigma)\delta({\bf r}_{ij})
   +t_1(1+x_1P_\sigma)\frac{1}{2\hbar^2}\{p_{ij}^2\delta({{\bf r}_{ij})}
      +h.c.\}\nonumber\\
& &  +t_2(1+x_2P_{\sigma})\frac{1}{\hbar^2}{\bf p}_{ij}.\delta({\bf r}_{ij})
   {\bf p}_{ij}
    +\frac{1}{6}t_3(1+x_3P_\sigma)\rho^\alpha\delta({\bf r}_{ij})\nonumber\\
& &    +\frac{i}{\hbar^2}W_0(\mbox{\boldmath$\sigma_i+\sigma_j$})
      {\bf .p}_{ij}\times\delta({\bf r}_{ij}){\bf p}_{ij}  \quad .
\eeqy

The total nuclear energy density at any point can now be written as  
\beqy\label{3.10}
{\mathcal E}_{nuc} = \sum_q\Big(\frac{\hbar^2}{2M^*_q}\tau_q \Big)
+ {\mathcal V}        \quad ,
\eeqy
where $\tau_q$ gives the kinetic-energy density of nucleons $q$ as 
$\frac{\hbar^2}{2M_q}\tau_q$ (the first term of Eq. ({3.10}) is the 
kinetic-energy density multiplied by $M_q/M^*_q$), the effective mass $M^*_q$ 
is given by 
\beqy\label{3.11}
\frac{\hbar^2}{2M^*_q}  & = &
\frac{\hbar^2}{2M_q}+\frac{1}{8}\Big\{t_1(2+ x_1)+
t_2(2+x_2)\Big \}\rho  \nonumber  \\
& & +\frac{1}{8}\Big\{t_2(1+2x_2)
-t_1(1+2x_1)\Big \}\rho_q                    \quad ,
\eeqy
and the static part of the potential energy by
\beqy\label{3.12}
{\mathcal V}
 & = & \frac{1}{4}t_0\Big \{(2+x_0)\rho^2-(1+2x_0)
            \sum_q\rho_q^2\Big \}  \nonumber\\
& & +\frac{1}{32}\Big \{3t_1(2+x_1)-t_2(2+x_2)\Big \}
       (\mbox{\boldmath$\nabla$}\rho)^2\nonumber\\
   & &-\frac{1}{32}\Big \{3t_1(1+2x_1)+t_2(1+2x_2)\Big \}
      \sum_q(\mbox{\boldmath$\nabla$}\rho_q)^2 \nonumber\\
   &&  +\frac{1}{24}t_3\Big \{ (2+x_3)\rho^2-(1+2x_3)
       \sum_q\rho_q^2 \Big \}\rho^\alpha \nonumber\\
  & & + \frac{1}{2}W_0\sum_q\mbox{\boldmath${\bf J}_q \cdot $}
\mbox{\boldmath$\nabla$}(\rho+\rho_q)  \quad ,
\eeqy
in which ${\bf J}_q$ is the spin-current density, and we have now set the 
``quadratic
current" term ${\mathcal V_{JJ}}$ of Eq. (A4c) of Ref. \cite{opp97} equal to 0 
throughout the calculation (our treatment of this term in Ref. \cite{opp97} was
inconsistent). For the nuclear free-energy density at any point we then have
\beqy\label{3.13}
{\mathcal F}_{nuc} = \sum_q {\mathcal K_q}  + {\mathcal V} \quad ,
 \eeqy
where 
\beqy\label{3.14}
{\mathcal K_q} = \frac{\hbar^2}{2M^*_q}\tau_q - T\sigma_q \quad .
\eeqy

The first stage of the full TETFSI method that we adopt in this paper for
a given temperature $T$, mean density $\bar{\rho}$ and fixed values of $N$ and
$Z$ consists in approximating the exact HF value of the nuclear free-energy 
density ${\mathcal F}_{nuc}$ for the given Skyrme force by the full 
fourth-order TETF method of BBD \cite{bbd}: see, for example, our paper 
\cite{opp97}, the appendix of which contains a convenient summary of the 
formalism, as we have used it here. We do not repeat this formalism here, 
although in Appendix \ref{cv} we present the TETF expansion for the specific 
heat, which appears not to have previously been published. Moreover, we remark 
here
that the TETF expression for $\sigma_q$ given by BBD \cite{bbd} assumes that 
\beqy\label{3.15}
\sigma = -\Big(\frac{\partial {\mathcal F}}{\partial T}\Big)_{\rho,X} \quad ,
\eeqy
where it is to be noted that it is the local density $\rho$, and not the mean
density $\bar{\rho}$, that is held constant.
We show in Appendix \ref{theorem} that Eq. (\ref{3.15}) follows from
Eq. (\ref{1.4}) only if equilibrium holds (this condition is made necessary
by virtue of the temperature dependence of the density distribution itself).

The essence of the TETF method is to express ${\tau_q}, {\bf J}_q$ and 
$\sigma_q$ in terms of an assumed density distribution, which here we take to 
have the form given in Eqns. (\ref{2.4}) and (\ref{2.5}). The value that we 
obtain for the total free energy per nucleon $f$, as given by Eqns. (\ref{3.3})
and (\ref{3.6}), is minimized with respect to the six geometrical 
parameters $\rho_{0q}$, $C_q$, $a_q$ of the parametrized nucleon distribution. 
(In looking for full nuclear and beta equilibrium we minimize with respect to 
$N$ and $Z$ also.) The resulting nucleon distributions are denoted by 
$\widetilde{\rho_q}$, and the corresponding approximations to ${\tau_q}, 
{\bf J}_q, \sigma_q$ and $f$ by $\widetilde{\tau_q}, \widetilde{{\bf J}_q},
\widetilde{\sigma_q}$ and $f_{TETF}$ respectively. The value of all these 
approximations to the exact HF values vary smoothly with respect to $N$ and 
$Z$: it is a characteristic of the TETF method that shell corrections are lost.
In the second stage we use the Strutinsky-integral method to correct $f_{TETF}$
perturbatively for proton shell effects, as follows.

At zero temperature, where $f_{TETF} = e_{ETF}$, the 
corrected value of $e$ takes the form 
\beqy\label{3.16}
e = e_{ETF} + \frac{1}{A}E^{sc}_p   \quad ,
\eeqy
where, according to the Strutinsky-integral theorem,
\beqy\label{3.17}
E^{sc}_p = \sum\limits_{i}n_i\widetilde{\epsilon_{i,p}} -
\int d^3{\bf r}\Big(\frac{\hbar^2}{2\widetilde{M^*_p}}\widetilde{\tau_p} + 
\widetilde{\rho_p}\widetilde{U_p}
+\widetilde{{\bf J}_p}\cdot\widetilde{{\bf W}_p}\Big) \quad .
\eeqy
Here the integral goes over the volume of the WS cell, while the sum goes over 
all the occupied s.p. proton states, with the s.p. energies 
$\widetilde{\epsilon_{i,p}}$ being the eigenvalues of the s.p. Schr\"odinger 
equation
\beqy\label{3.18}
\left\{-\bfdel\frac{\hbar^2}{2\widetilde{M^*_p}({\bf r})}\cdot\bfdel + 
\widetilde{U_p}({\bf r}) - i\widetilde{{\bf W}_p}({\bf r})\cdot\bfdel
\times\mbox{\boldmath$ \sigma$}\right\}\phi_{i,p}=
\widetilde{\epsilon_{i,p}}\phi_{i,p} \quad ,
\eeqy
and $n_i$ the occupancy of s.p. state $i$ (= 0 or 1 for $T$ = 0 and
no pairing).
In these last two equations the effective mass $\widetilde{M^*_p}$ is given
by Eq. (\ref{3.11}) with the smooth ETF densities $\widetilde{\rho_q}$
replacing the exact HF densities $\rho_q$. Likewise $\widetilde{U_p}$ and 
$\widetilde{{\bf W}_p}$ are the central and spin-orbit proton fields, 
respectively, given by the usual HF expressions for these fields (see, for 
example, Eqns. (7) and (9) of Ref. \cite{far01}), with $\widetilde{\rho_q}$
replacing $\rho_q$ (note that here $\widetilde{U_p}$ contains the Coulomb 
field). 
      
All three of these fields involve a folding of the Skyrme force over the 
nucleon distribution $\widetilde{\rho_q}$ that emerges from the minimisation of
$e_{ETF}$ in the ETF part of the calculation. The fact that the same Skyrme 
force underlies also the ETF part of the calculation implies a high degree of 
consistency between the two parts, which probably accounts for the close 
agreement found in comparisons with exact self-consistent HF calculations 
\cite{dut,ton} (note that these tests were limited to the case of bound nuclei 
at $T = 0$ and with $M^*_q = M_q$). This theorem seems to have been stated for 
the first time in Ref. \cite{cjb77}. A derivation, limited, however, to 
the case $M_q^* = M_q$, was sketched in Ref. \cite{abo2}. A more complete 
proof, applicable to arbitrary effective mass, is presented in Appendix
\ref{etfsi}.

As for $T > 0$, in Ref. \cite{dop04} we replaced Eq. (\ref{3.16}) by
\beqy\label{3.19}
f = f_{TETF} + \frac{1}{A}E^{sc}_p   \quad ,
\eeqy
where the proton shell correction is still given by Eq. (\ref{3.17}), with
\beqy\label{3.21}
n_i = \frac{1}{1 + \exp\{(\widetilde{\epsilon_{i,p}} - \mu_p)/T\}} \quad ,
\eeqy
$\mu_p$ being the chemical potential for protons. However, this does not take 
account of possible proton shell effects in the entropy, so here we will write 
rather
\beqy\label{3.22}
f_{TETFSI} = f_{TETF} + \frac{1}{A}E^{sc}_p - T(s^{s.p.}_p - s^{TETF}_p)  
\quad ,
\eeqy
where $s^{s.p.}_p$ is the usual s.p. expression for the proton entropy,
\beqy\label{3.23}
s^{s.p.}_p = -\sum\limits_{i}\left\{n_i\ln\,n_i + 
(1 - n_i)\ln\,(1 - n_i)\right\} \quad ,
\eeqy
in which the sum goes over all proton states.

To determine the eigenvalues $\widetilde{\epsilon_{i,p}}$ we expand the 
eigensolutions
$\phi_{i,p}$ to Eq. (\ref{3.18}) in the basis defined by spherical Bessel
functions $j_l(k_nr)$ with the $k_n$ chosen to satisfy homogeneous 
boundary conditions (vanishing of the function or of its first derivative) on 
the surface of the WS cell.
This generates a complete set of functions that are orthogonal over the cell, 
and we diagonalize the associated matrix. 

{\it Interpolation Schemes.} In addition to its much greater simplicity and
rapidity, as compared to full-blown HF calculations, the (T)ETFSI method has
the advantage of lending itself to interpolation. The point is that while the
shell corrections are indeed fluctuating quantities, the method expresses
these quantities in terms of quantities that themselves vary smoothly: the 
fluctuations
arise entirely in the summation indicated on the right-hand side of 
Eq. (\ref{3.17}). This feature was heavily exploited in the construction of the
ETFSI-1 mass table, the first mass table to be based on microscopic forces 
\cite{abo2}, and we anticipate that it could prove equally fruitful for the
extensive tabulation of the EOS described in this paper.

{\it Specific heat.} It is convenient to define a density of specific heat 
at constant volume, ${\mathcal C_v}({\bf r})$, at each point in the cell, 
according to
\beqy\label{3.25}
c_v = \frac{1}{A}\int{\mathcal C_v}({\bf r})d^3{\bf r} \quad ,
\eeqy
where $c_v$ is the specific heat per nucleon (\ref{1.3}), and the integration 
goes over the volume of the WS cell. It then follows,
provided equilibrium holds, that
\beqy\label{3.26}
{\mathcal C_v}({\bf r}) = 
T\Big(\frac{\partial \sigma}{\partial T}\Big)_{\rho,X}
\eeqy
(see Appendix \ref{theorem}). This last result can then be used, starting 
from the
TETF expansion of $\sigma$ \cite{bbd}, to derive the corresponding expansion 
for ${\mathcal C_v}$ up to order $\hbar^4$. This expansion is presented in 
Appendix \ref{cv}.

{\it Convergence of the TETF expansions.}

Before applying the TETFSI method to the calculation of the neutron-star crust 
we make extensive tests of the convergence of the series expansion implicit in 
the TETF method. For this it is convenient to define the quantity
\beqy\label{3.27}
\phi_q = \frac{4\pi}{A}\int_0^{R_c}{\mathcal K}_q(r)\,r^2\,dr \quad  ,
\eeqy
which is just the ``kinetic-thermal" part of the nuclear free energy per
nucleon. 

We consider two densities, $\bar{\rho}$ = 0.06 and 3 $\times 10^{-4}$
nucleons.fm$^{-3}$, which lie close to the upper and lower limits, 
respectively, of the density range encountered in the inner crust. For the 
former we take a cell with $Z = 30, A$ = 1123, and for the latter $Z = 40,
A = 150$, these lying close to the equilibrium configuration (see Section
\ref{equilib}). The results are shown in Tables \ref{tab1} and \ref{tab2},
respectively, for three temperatures, 0.1, 1.0, and 3.0 MeV. For each quantity,
$\phi_n$ and $\phi_p$, we show in successive lines the Thomas-Fermi  (TF)
approximation, the second-order correction in $\hbar$, and the fourth-order
correction; the sum of these three contributions is shown in the adjacent 
column. We also show in these tables the corresponding expansion terms for
$s_n, s_p, c_{v,n},c_{v,p}$ and the proton entropy as
calculated by the s.p. expression (\ref{3.23}).  

We see that the expansions for all the neutron quantities,
$\phi_n, \phi_p, s_n,$ and $c_{v,n},$ 
converge well over the entire density and temperature range. However, the 
expansions for the proton entropy $s_p$ and the related specific heat $c_{v,p}$
diverge badly at low temperature, to the point of occasionally giving negative
entropies and specific heats. This proton-related problem is seen to be 
worse at low densities, which suggests that the fact that the neutron-related 
quantities converge so much better might be related to the fact that our 
neutron densities are always much higher than the proton densities. Indeed, we
find that at an outer-crust density of $\bar{\rho}$ = 1 $\times 10^{-4}$
nucleons.fm$^{-3}$ the TETF expansion for the neutron entropy of $^{208}$Pb 
converges rather poorly. The existence of a significant background term, which
gives no contribution to either the second- or fourth-order terms, appears to 
be crucial in this respect. It is fortunate that the lower density limit
for good convergence of the TETF expansion for the neutron entropy lies below
our domain of interest. In any case, it is quite clear that we cannot use the 
TETF expansions for the proton entropy or specific heat. 

For the proton entropy we therefore fall back instead on the s.p. expression 
(\ref{3.23}), which we require anyway for the proton shell corrections. On the 
other hand, we recall that we cannot use this s.p. expression for the neutron 
entropy because of continuum problems: it was precisely for this reason that 
we had to abandon the HF approach. As for the proton specific heat, we could in
principle calculate numerically the temperature derivative of the proton 
entropy (\ref{3.23}) and use Eq. (\ref{1.3}), but this is too time-consuming, 
if done accurately. We thus simply take the TF approximation for the proton
contribution to the specific heat. Since the neutron contribution dominates 
the specific heat, the error thereby introduced will be relatively small;  
in any case we see that at higher temperatures the TF approximation to the 
entropy agrees approximately with the entropy calculated by Eq. (\ref{3.23}).
   
It is reasonable to ask why the TETF entropy expansion diverges at low T.
The second-order term contains a $T^{-1}$ factor, and the fourth-order term a
$T^{-2}$ factor (see Eqns. (A.25b-c) of Ref. \cite{opp97}), so that the 
corresponding numerators must likewise vanish at low $T$. This suggests that
a massive cancellation within these numerators is leading to a significant loss
of precision. However, we derived the low-$T$ (strong degeneracy) limit of the
TETF entropy expansion (see Appendix \ref{degen}), and found essentially the 
same
numerical results. We suggest that the reason for the observed breakdown in
the TETF expansions of the entropy and specific heat lies with the fact that
the validity of Eqns. (\ref{3.15}) and (\ref{3.26}) requires that the system
be in equilibrium, which can never be exactly the case when the density 
profiles are parametrized, as in Eqns. (\ref{2.4}) and (\ref{2.5}).

\section{Equilibrium properties of inner crust}
\renewcommand{\theequation}{4.\arabic{equation}}
\setcounter{equation}{0}
\label{equilib}

\begin{figure}
\centerline{\epsfig{figure=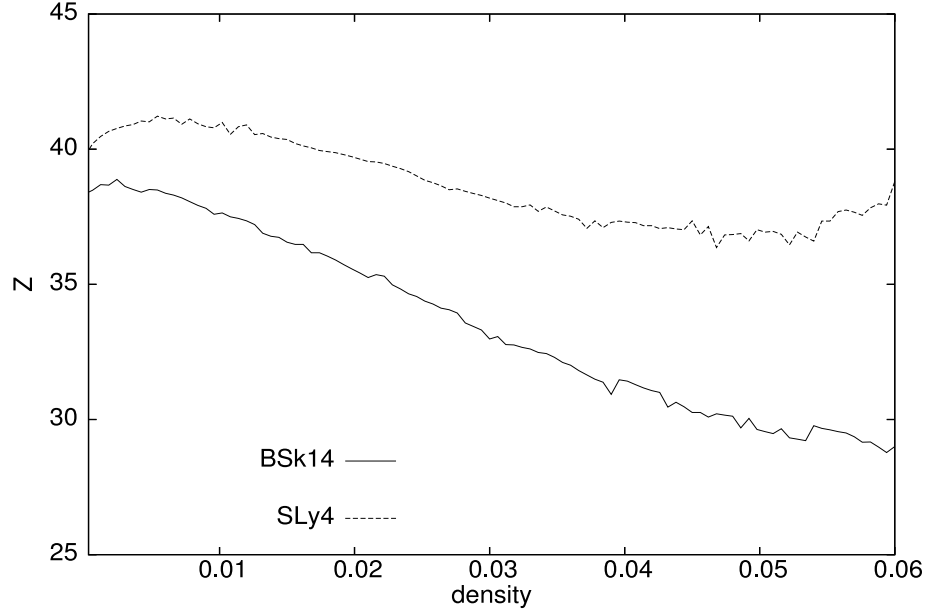,height=13.0cm,width=9cm,angle=-90}}
\caption{Number of protons $Z$ in WS cell given by ETF method for nuclear and 
beta equilibrium at $T = 0$ as a function of density (fm$^{-3}$) for forces 
BSK14 and SLy4.}
\label{fig1}
\end{figure}

\begin{figure}
\centerline{\epsfig{figure=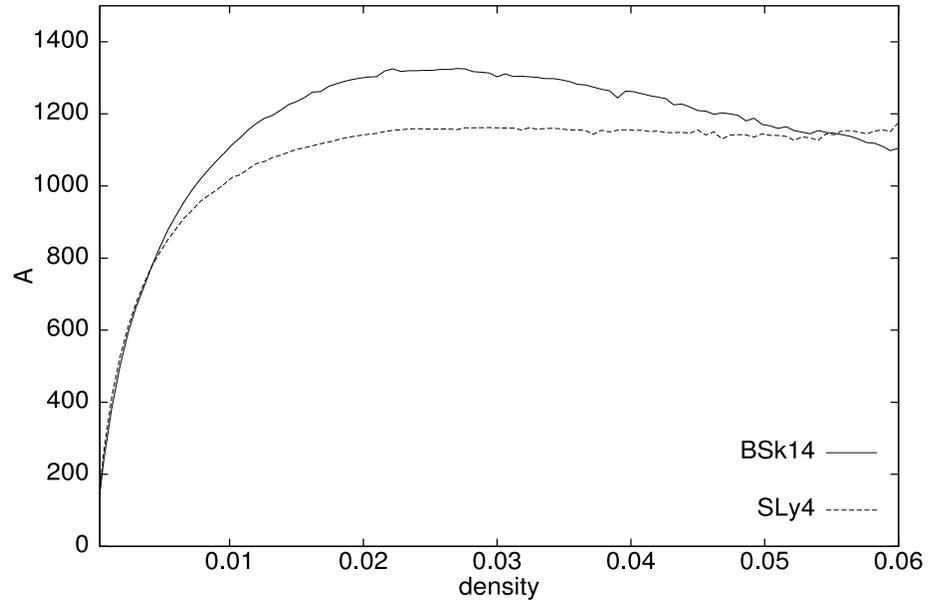,height=13.0cm,width=9cm,angle=-90}}
\caption{Total number of nucleons $A = Z + N$ in WS cell given by ETF method 
for nuclear and beta equilibrium at $T = 0$ as a function of density 
(fm$^{-3}$) for forces BSK14 and SLy4.} 
\label{fig2}
\end{figure}

\begin{figure}
\centerline{\epsfig{figure=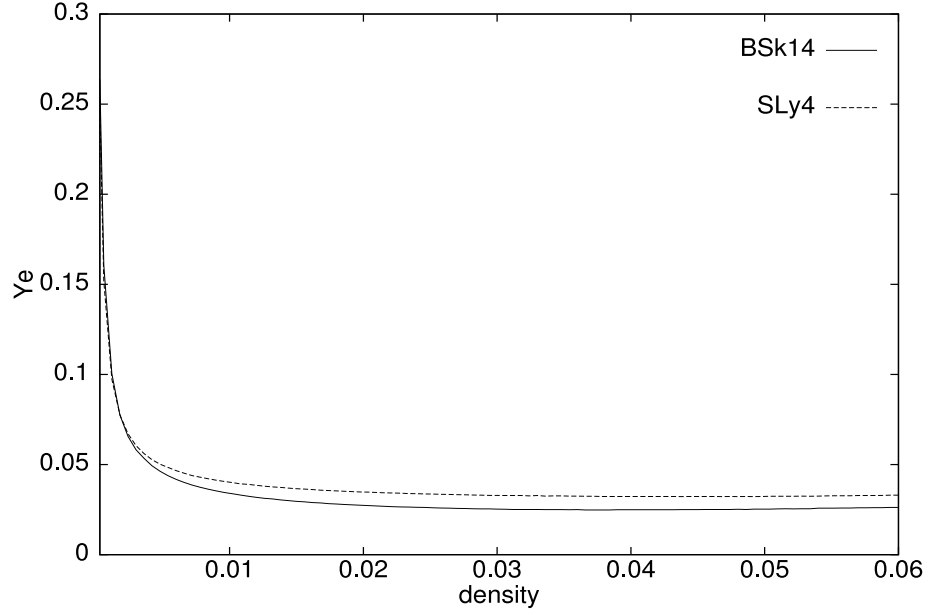,height=13.0cm,width=9cm,angle=-90}}
\caption{Fraction $Y_e = Z/A$ of nucleons that are protons given by ETF method 
for nuclear and beta equilibrium at $T = 0$ as a function of density 
(fm$^{-3}$) for forces BSK14 and SLy4.}
\label{fig3}
\end{figure}

{\it Pure TETF calculations.} In calculating the equilibrium (nuclear and beta) 
composition of the inner crust, i.e., the number 
$Z$ of protons and the number $N$ of neutrons per WS cell at any given mean 
density $\bar{\rho}$, we shall first neglect shell corrections. For this we 
simply minimize $f_{TETF}$, calculated at $T = 0$, with respect to $N$ and $Z$.
The results for $Z, A = Z + N$, and $Y_e = Z/A$ are shown in Figs. \ref{fig1},
\ref{fig2}, and \ref{fig3}, respectively. We stress that in this picture $Z$
and $A$ vary continuously, and are not restricted to integral values. 

These figures also show the values of the same quantities that we find for the 
force SLy4 \cite{cha98}, which has been widely applied to neutron stars. 
Fig. \ref{fig2} for $A$ shows that at a given $\bar{\rho}$ the WS cells tend to
be bigger for force BSk14, presumably because the value of the surface-energy
coefficient $a_{sf}$ is slightly larger for the former (18.11 MeV as opposed
to 17.6 MeV). However,
since $Y_e$ runs slightly higher for SLy4 (see Fig. \ref{fig3}) there are
generally somewhat more protons for this latter force, as seen in Fig. 
\ref{fig1}. (The higher values of $Y_e$ found for SLy4 can be traced to the 
fact that in homogeneous neutron matter the energy per neutron is higher for
SLy4 over the entire density range considered here.) This sensitivity to the 
choice of force will have implications
both for transport properties in the inner crust and for nucleosynthesis in 
decompressing neutron-star matter.

\begin{figure}
\centerline{\epsfig{figure=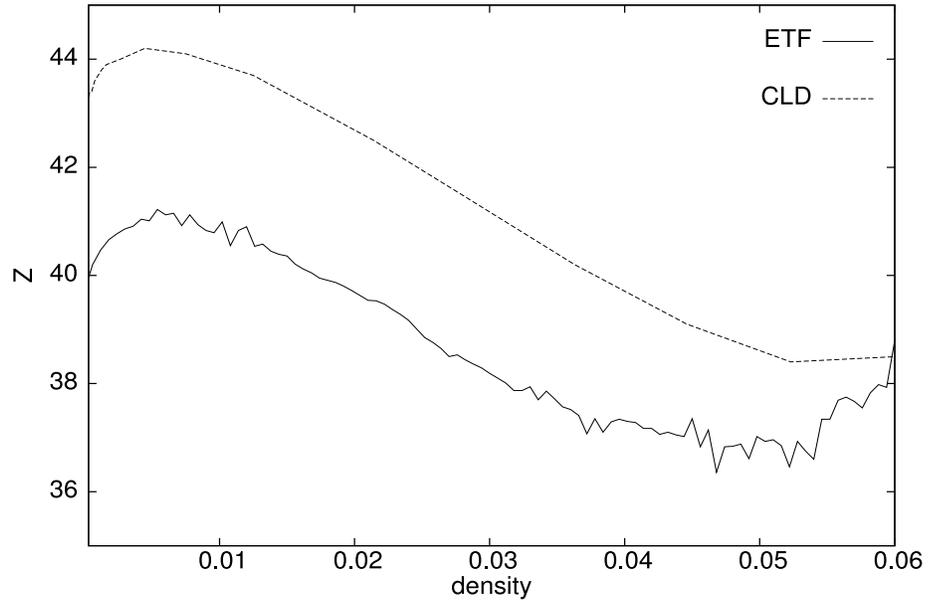,height=13.0cm,width=9cm,angle=-90}}
\caption{Comparison of ETF and CLD (compressible liquid drop) calculations
of equilibrium value of $Z$ at $T = 0$ as a function of density
(fm$^{-3}$) with SLy4 force.} 
\label{fig4}
\end{figure}

Now the popular EOS of Ref. \cite{dh01} is also based on the SLy4 force, to the
extent that the parameters of the underlying compressible liquid-drop model are
calculated for this force. We compare the results of Ref. \cite{dh01} for $Z$ 
with our own in Fig. \ref{fig4}, where it will be seen that the two models lead
to appreciable differences even before taking shell effects into account.  

{\it TETFSI calculations: proton shell effects.} When shell corrections are 
included $Z$ and $A$ must be considered as integers,
and thus change discontinuously: each $(Z, A)$ pair defines a phase. Since 
$\bar{\rho}$ changes continuously transitions between one phase and another 
will take place over a finite range of $\bar{\rho}$, in which the two phases
coexist. We shall neglect this feature (see, however, the comments below),
but our results will be valid over those considerable 
intervals of $\bar{\rho}$ for which only a single phase exists.

The calculations proceed as outlined in Section \ref{method}. Thus for given
values of $T$ and $\bar{\rho}$, and for a given $(Z, A)$ pair, $f_{TETF}$ has 
to be minimized with respect to the geometrical parameters of the cell, and
then shell-corrected according to Eq. (\ref{3.19}) (note that the (T)ETFSI
method calculates shell corrections perturbatively). The equilibrium values
of $Z$ and $A$, i.e., the values that minimize the TETFSI value of $f$ for 
given values of $T$ and $\bar{\rho}$, are shown as functions of $\bar{\rho}$
for $T$ = 0, 0.1, and 1.0 MeV in Tables \ref{tab3}, \ref{tab4}, and \ref{tab5},
respectively. 

\begin{figure}
\centerline{\epsfig{figure=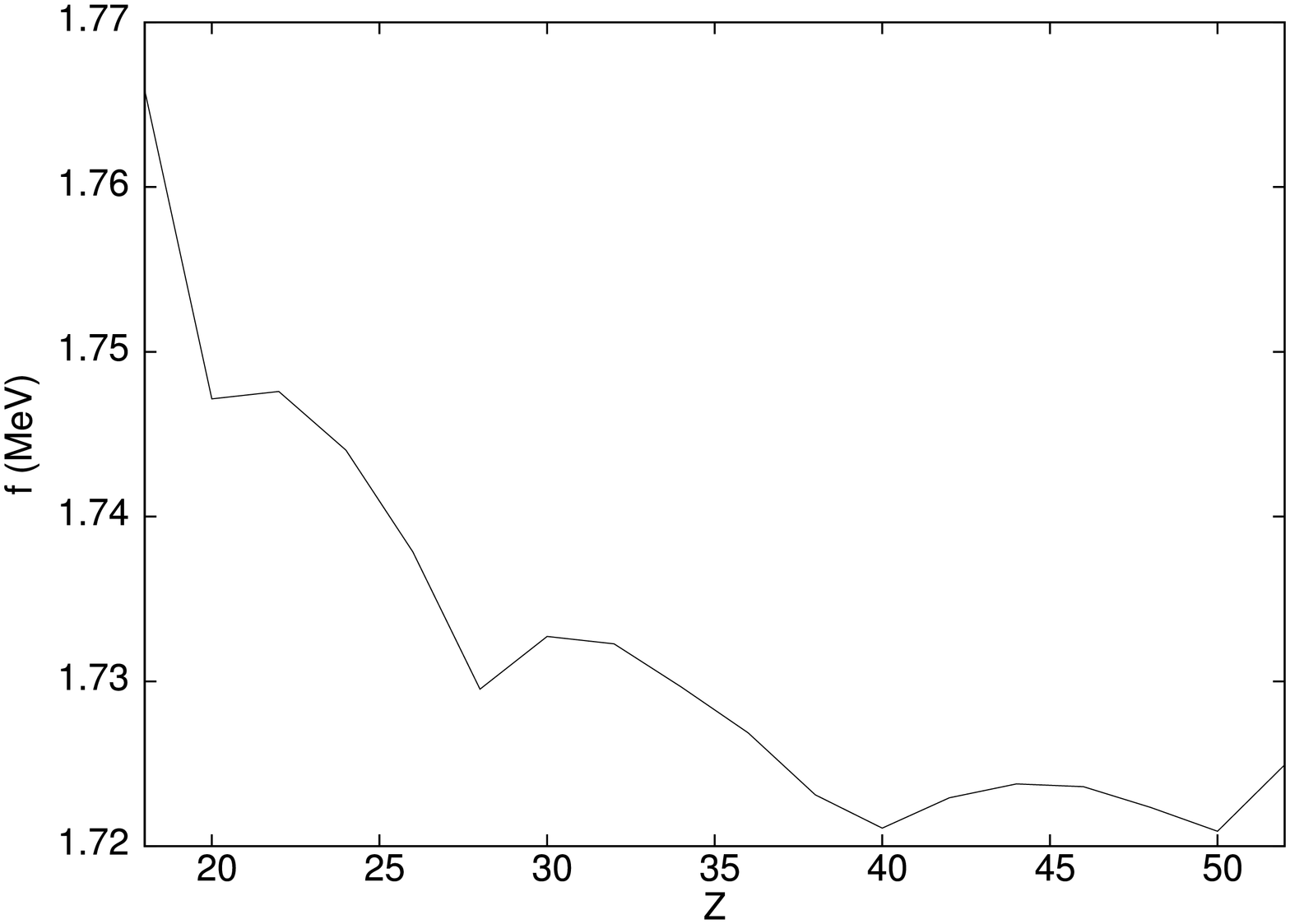,height=13.0cm,width=9cm,angle=-90}}
\caption{Variation of $f_{TETFSI}$ with $Z$ (always for optimal value of $A$)
at $\bar{\rho}$ = 0.005 fm$^{-3}$ and $T$ = 0.1 MeV.}
\label{fig5}
\end{figure}

\begin{figure}
\centerline{\epsfig{figure=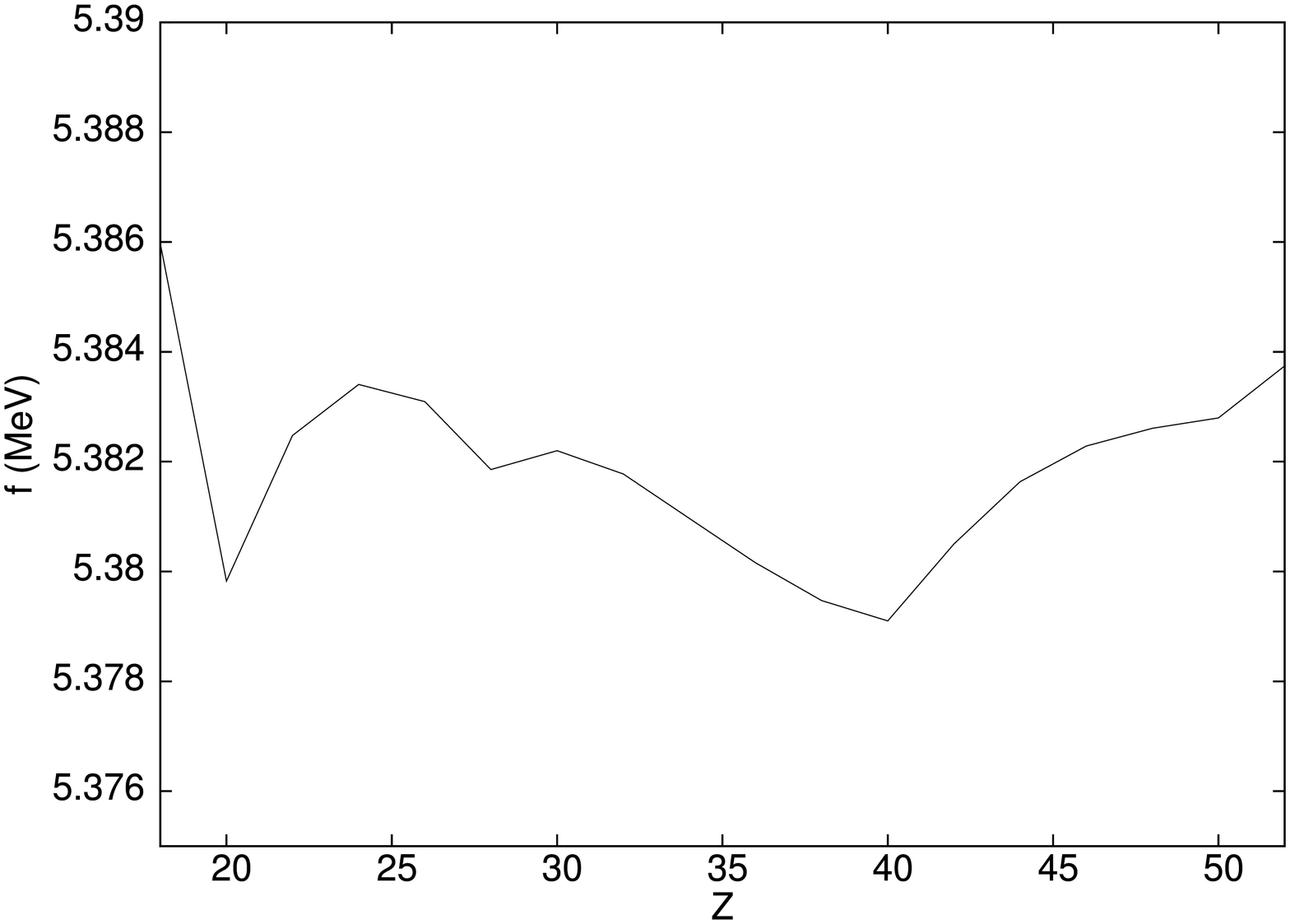,height=13.0cm,width=9cm,angle=-90}}
\caption{Variation of $f_{TETFSI}$ with $Z$ (always for optimal value of $A$)
at $\bar{\rho}$ = 0.04 fm$^{-3}$. and $T$ = 0.1 MeV.}
\label{fig6}
\end{figure}

With the TETF values being shown in parentheses, we see in Tables \ref{tab3}
and \ref{tab4} that at $T$ = 0 and 0.1 MeV there are strong proton shell 
effects, with $Z$ = 50, 40, and 20 being successively favored as the density 
increases; in fact, these are the {\it only} values of $Z$ that appear. 
The changes 
from $Z$ = 50 at $\bar{\rho}$ = 0.005 fm$^{-3}$ to $Z$ = 40 at 0.01 
fm$^{-3}$ and from $Z$ = 40 at $\bar{\rho}$ = 0.04 fm$^{-3}$ to $Z$ = 20
at 0.05 fm$^{-3}$ appear to be discontinuous. This situation can easily be 
understood from Figs. \ref{fig5} and \ref{fig6}, where we show $f$ at $T$ = 
0.1 MeV as a function of $Z$ for $\bar{\rho}$ = 0.005 and 0.04 fm$^{-3}$, 
respectively, minimizing with respect to $A$ for each value of $Z$: 
relatively strong minima occur for $Z$ = 50, 40 and 20, and the system flips 
from one to the next as the density increases. (Somewhat surprisingly, although
$Z$ = 28 is consistently a strong local minimum it is never an absolute 
minimum.) It is noteworthy that the value of $Z$ = 50
that we find here at the outside edge of
the inner crust agrees with what we find for an {\it outer}-crust calculation
using the HFB-14 mass model \cite{gsp07}, i.e., a mass model based on the
BSk14 force used here.

It will be seen from Figs. \ref{fig5} and \ref{fig6} that the minima are very
close in energy, and numerical uncertainties in our computation often make it 
impossible to affirm with certainty which magic number prevails at a given
density. Certainly, changing the force could be expected to lead to changes
in the sequence of magic numbers. In fact, Negele and Vautherin \cite{nv73} 
report a quite difference sequence for the proton numbers 40 and 50 (and do
not find $Z$ = 20 at any density).  
At $T$ = 0.1 MeV the energy fluctuations associated
with the Boltzmann factor represent an uncertainty on the total cell energy, 
and thus amount to 0.1/$A$ MeV per nucleon. Reference to Figs. \ref{fig5} and 
\ref{fig6} then shows that there will be negligible admixture of other values
of $Z$ with the magic values, although there could be significant
admixture of $A$-values. 

Tables \ref{tab3} and \ref{tab4} reveal no essential difference between 
$T$ = 0 and $T$ = 0.1 MeV. However, with the energy differences between 
adjacent magic numbers being so small it could be that there are some 
intermediate densities for which the composition changes as $T$ varies
between 0 and 0.1 MeV. This is one sense in which caution might have to be
exercised in adopting the picture of ``cold catalyzed matter".

Table \ref{tab5} shows that at $T$ = 1 MeV shell effects have effectively been 
wiped out, even though there are still significant differences between the 
TETF and TETFSI values. This raises the question of whether a nuclear 
equilibrium in the crust of a cooling neutron star can be maintained right
down to $T$ = 0.1 MeV. If ``freeze-out" were to occur at, or slightly below,
$T$ = 1 MeV, i.e., if the complex rearrangement of nucleons necessary to
maintain nuclear equilibrium could no longer take place over the lifetime of
the neutron star, then the sharp shell effects that we have predicted here
would not be observed. This is another, and probably more serious sense, in
which the picture of ``cold catalyzed matter" has to be carefully scrutinized. 

{\it Pressure.} We extract $P$ directly from the computed values of 
$f_{TETFSI}$ using Eq. (\ref{1.2}), numerically evaluating the derivative
with a 3-point Savitzky-Golay filter (routine `savgol' \cite{numrec}). Our
results are shown in the penultimate columns of Tables \ref{tab3}, \ref{tab4}, 
and \ref{tab5}; tests show that over this density range our results for $P$ are
reliable to within about 1 \%. 

Another quantity of astrophysical interest is the temperature variation of
the pressure $\Big(\frac{\partial P}{\partial T }\Big)_{\bar{\rho}}$; using a 
well known Maxwell relation we have
\beqy\label{4.1}
\Big(\frac{\partial P}{\partial T }\Big)_{\bar{\rho}} =
-\bar{\rho}^2\Big(\frac{\partial s}{\partial  \bar{\rho}}\Big)_T \quad.
\eeqy
This too requires a numerical differentiation, but only one: without using
the Maxwell relation we would have to perform {\it two} numerical
differentiations, with consequent loss of precision. Moreover, the derivative
of $s$ can be computed simultaneously with the one of $f$ that gives us $P$,
with negligible increase in computer time. The results are shown in the last 
columns of Tables \ref{tab3}, \ref{tab4}, and \ref{tab5}. 

{\it Phase equilibrium.} If we were to consider a quasi-continuum of values of 
${\bar{\rho}}$ our calculations might show unphysical discontinuities in the 
pressure at the transition between the different ($Z, A$) phases.
This is a result of our neglect of the possibility of a thermal equilibrium 
between the two phases in question, and in reality the pressure remains
continuous. The transition pressure $P$ is characterized by equality of the 
Gibbs free energy per nucleon in each phase,
\beqy\label{4.2}
g_1 = g_2 \quad  ,
\eeqy
where
\beqy\label{4.3}
g_i = f_i + \frac{P}{{\bar{\rho}}_i} \quad . 
\eeqy
We stress that the equilibrium pressure satisfying this condition can be 
determined by calculating $f_i$ as a function of ${\bar{\rho}}$ for each phase
separately; in particular it is at no point necessary to minimize the Gibbs
function itself. Moreover, it can be shown that the condition (\ref{4.2}) 
follows from a minimization of the total {\it Helmholtz} function \cite{adk}. 

\section{Conclusions}

We have developed here a high-speed approximation to the HF method for 
calculating the EOS of the neutron-star inner crust with Skyrme-type forces. 
Our method, which we refer to as the TETFSI method, models the inner
crust in terms of the Wigner-Seitz cell, and consists essentially of a 
generalization to finite temperatures (and arbitrary effective mass) of the
ETFSI method originally developed as a mass model \cite{dut,ton,pea,abo1,abo2}.
An essential difference between our TETFSI method and a full-scale HF
calculation of the EOS is that, whereas the latter method inevitably and 
automatically calculates both neutron and proton shell effects, here we 
calculate only the latter, since in reality shell effects are much weaker for 
neutrons than for protons, and will have negligible impact on the composition.
In fact, if the HF method is used in a WS picture, as in the classical work of 
Negele and Vautherin\cite{nv73} it will lead, because of
discretization, to spuriously large neutron shell effects\cite{cha07}. 
As in Ref. \cite{nv73}, we have neglected pairing in this paper, pending the 
determination of an effective pairing interaction appropriate to the conditions
pertaining in neutron-star crusts. Nevertheless, it will be easy to include 
pairing in the (T)ETFSI framework, as already done in the 
ETFSI mass models \cite{dut,ton,pea,abo1,abo2}. 

It was found that the TETF expansion of the entropy converges badly at low
temperatures for densities typical of inner-crust protons (there was no problem
for neutrons). We solved this  difficulty by using the s.p. expression for the
proton entropy. 

Our exploratory calculations of the EOS were performed with the Skyrme-type
force BSk14, a force that was fitted to essentially all the nuclear-mass data, 
forming thereby the basis of the HFB-14 mass model \cite{gsp07}. This force
is particularly suitable for calculating the properties of neutron-star crustal
matter, because it has been fitted to the properties of homogeneous neutron 
matter while at the same time the good fit to masses ensures that both
inhomogeneities and the neutron-proton interaction are well represented.

The calculated composition of the WS cells representing the clustering in the
inner crust showed striking shell effects: for $T$ = 0 
the proton number $Z$ was limited to the magic 
values of 50, 40 and 20, the value decreasing with increasing density (at the
interface with the outer crust we found continuity with an outer-crust
calculation based on the HFB-14 mass model). Although essentially identical 
results are obtained for $T$ = 0.1 MeV, all our calculated shell effects are
wiped out at $T$ = 1 MeV, which means that whether or not shell effects
actually exist in the cold crust depends very much on the ``freeze-out"
temperature for nuclear equilibrium. On the other hand, we have shown that even
without taking shell effects into account there are considerable differences
between our predictions and those of the compressible liquid-drop model on
which the EOS of Ref. \cite{dh01} is based. 

We intend to apply the method described here to a study of the synthesis of
r-process nuclei in decompressing neutron-star crustal matter. To this end
we present here, apparently for the first time, the TETF expressions for
the specific heat of an inhomogeneous system of nucleons. In this same context
of extensive computations over a wide range of temperature, density and 
composition, we point out that the (T)ETFSI method lends itself admirably to
interpolation, without any loss of precision in the calculated shell effects,
essentially because these arise in the sums of quantities that themselves
vary smoothly \cite{pea}.

\begin{acknowledgments}

We are indebted to B. K. Jennings for suggesting to us the Strutinsky-integral 
method. M. Brack and J. M. Lattimer are thanked for stimulating discussions 
in the early stages of this work. The financial support of the NSERC (Canada) 
and the FNRS (Belgium) is acknowledged. N.C. gratefully acknowledges the award
of a Marie Curie Intra-European fellowship (contract number 
MEIF-CT-2005-024660).
\end{acknowledgments}

\appendix

\section{TETF expressions for the specific-heat density ${\mathcal C}_v$}
\label{cv}
\renewcommand{\theequation}{A.\arabic{equation}}
\setcounter{equation}{0}

We expand the specific-heat density ${\mathcal C}_v$ of Eq. (\ref{3.26})
according to

\beqy\label{cv.0}
{\mathcal C}_v = {\mathcal C}_v^{(TF)} + {\mathcal C}_v^{(2)} +
{\mathcal C}_v^{(4)} + \cdots  \quad ,
\eeqy
where the first term on the right-hand side represents the Thomas-Fermi
approximation, the second term the first-order correction in $\hbar^2$, and the
last term the second-order correction in $\hbar^2$. Then

\bmlet
\beqy\label{cv.1a}
{\mathcal C}^{(TF)}_{v,q} = \frac{3}{2}\left\{\sigma^{(TF)}_q + \eta_q\rho_q
-3\rho_q\frac{I_{1/2}(\eta_q)}{I_{-1/2}(\eta_q)}\right\}
\eeqy
\beqy\label{cv.1b}
{\mathcal C}^{(2)}_{v,q} = -\frac{\sigma^{(2)}_q}{16\nu_q}(24x^3 + 22x^2 + 5x - 63xy
- 33y +45z)
\eeqy
\beqy\label{cv.1c}
{\mathcal C}^{(4)}_{v,q} = 
-2\sigma^{(4)}_q - 3\left(\frac{\hbar^2}{2MT}\right)^2
\frac{I_{1/2}(\eta_q)}{I_{-1/2}(\eta_q)}
\sum_{i=1,3}G_i^q\left(\frac{\partial\chi_i}{\partial\eta_q}\right)_T
\eeqy
\emlet

In Eq. (\ref{cv.1c}) we have 
\bmlet
\beqy\label{cv.2a}
G_1^q = \frac{(\bfdel^2\rho_q)^2}{\rho_q}
\eeqy
\beqy\label{cv.2b}
G_2^q = \frac{\bfdel^2\rho_q(\bfdel\rho_q)^2}{\rho_q^2}
\eeqy
\beqy\label{cv.2c}
G_3^q = \frac{(\bfdel\rho_q)^4}{\rho_q^3}
\eeqy
\emlet
and
\bmlet
\beqy\label{cv.3a}
\left(\frac{\partial\chi_1}{\partial\eta_q}\right)_T &=& 
\left\{\frac{I_{-1/2}(\eta_q)}{I_{1/2}(\eta_q)}\right\}^2 
\Bigg(\frac{11}{192}x^3 - \frac{11}{60}xy - \frac{109}{320}x^2y 
+ \frac{5}{64}x^4 + \frac{25}{64}xz \nonumber \\
&+& \frac{5}{576}x^2 + \frac{11}{64}z
- \frac{1}{64}y + \frac{3}{20}y^2 - \frac{21}{64}w
\Bigg) 
\eeqy
\beqy\label{cv.3b}
\left(\frac{\partial\chi_2}{\partial\eta_q}\right)_T &=&
\left\{\frac{I_{-1/2}(\eta_q)}{I_{1/2}(\eta_q)}\right\}^2
\Bigg(\frac{85}{288}x^4 - \frac{119}{96}w + \frac{1}{6}z + \frac{187}{320}y^2
+ \frac{1}{18}x^3 - \frac{161}{80}x^3y \nonumber \\
&+& \frac{609}{320}xy^2 - \frac{175}{64}xw + \frac{159}{64}x^2z
- \frac{117}{64}yz -\frac{629}{480}x^2y - \frac{11}{60}xy
+ \frac{289}{192}xz  \nonumber \\
&+& \frac{35}{96}x^5 + \frac{63}{32}v\Bigg)
\eeqy
\beqy\label{cv.3c}
\left(\frac{\partial\chi_3}{\partial\eta_q}\right)_T &=&
\left\{\frac{I_{-1/2}(\eta_q)}{I_{1/2}(\eta_q)}\right\}^2
\Bigg(-\frac{77}{256}w + \frac{11}{192}x^4 + \frac{33}{256}y^2 - 
\frac{693}{256}u - \frac{391}{240}x^3y + \frac{2047}{1280}xy^2 \nonumber \\ 
&-& \frac{719}{320}x^4y - \frac{161}{64}xw + \frac{97}{32}x^3z - 
\frac{63}{16}x^2w + \frac{1071}{256}xv + \frac{69}{32}x^2z \nonumber \\
&-& \frac{207}{128}yz + \frac{315}{128}yw -\frac{11}{40}x^2y
+ \frac{11}{32}xz + \frac{161}{576}x^5 + \frac{483}{256}v \nonumber \\ 
&+& \frac{21}{64}x^6 + \frac{4317}{1280}x^2y^2 - \frac{645}{128}xyz
- \frac{801}{1280}y^3 + \frac{135}{128}z^2 \Bigg)   \quad .
\eeqy
\emlet
All quantities shown here are as defined in the Appendix of Ref. \cite{opp97}
and Appendix B of Ref. \cite{bbd}, except for $u$, which we define according to
\beqy\label{cv.4}
u = \frac{(I_{1/2})^6I_{-13/2}}{(I_{-1/2})^7} \quad .
\eeqy
Note that Eq. (\ref{cv.1c}) for the fourth-order term is valid only for an
effective mass $M_q^* = M_q$ (see the Appendix of Ref. \cite{opp97}).

As is the usual practice in expositions of the TETF method \cite{bbd}, we have 
dropped terms here that vanish on integrating over configuration space, which
in the present case corresponds to the WS cell. That is why we have required
the density gradients to vanish at the surface of the cell (see Section
\ref{wscell}).

\section{Proof of Eqns. (\ref{3.15}) and (\ref{3.26})}
\label{theorem}
\renewcommand{\theequation}{B.\arabic{equation}}
\setcounter{equation}{0}

The total free energy of the WS cell can be written as
\beqy\label{B.1}
F = \int{\mathcal F}(\rho, \bfdel\rho, T)d^3{\bf r}   \quad ,
\eeqy
where the integration goes over the volume of the cell. For the entropy of the
cell we have
\beqy\label{B.2}
S &=& -\Big(\frac{\partial F}{\partial T}\Big)_{\bar{\rho}} = 
-\int\Big(\frac{\partial {\mathcal F}}{\partial T}\Big)_{\bar{\rho}}
d^3{\bf r} \nonumber \\
&=& -\int\left\{\left(\frac{\partial {\mathcal F}}{\partial T}\right)_{\rho}
+ \left(\frac{\partial {\mathcal F}}{\partial \rho}\right)_T
\left(\frac{\partial \rho}{\partial T}\right)_{\bar{\rho}}
+ \left(\frac{\partial {\mathcal F}}{\partial {\bf \bfdel}\rho}\right)_T 
. \left(\frac{\partial {\bf \bfdel}\rho}{\partial T}\right)_{\bar{\rho}}
\right\}d^3{\bf r} \quad .
\eeqy
But, integrating by parts, we have
\beqy\label{B.3}
&&\int\left(\frac{\partial {\mathcal F}}{\partial \bfdel\rho}\right)_T . 
\left(\frac{\partial \bfdel\rho}{\partial T}\right)_{\bar{\rho}}d^3{\bf r} =
\int\left(\frac{\partial {\mathcal F}}{\partial \bfdel\rho}\right)_T .
\bfdel\left(\frac{\partial \rho}{\partial T}\right)_{\bar{\rho}}d^3{\bf r} 
\nonumber \\
&=&
-\int\left(\frac{\partial \rho}{\partial T}\right)_{\bar{\rho}} \bfdel
. \left(\frac{\partial {\mathcal F}}{\partial \bfdel\rho}\right)_T
d^3{\bf r}   \quad ,
\eeqy
where we are making use of the vanishing of 
$\left(\partial {\mathcal F}/\partial \bfdel\rho\right)_T$ on the 
surface of the cell; this follows from the fact that ${\mathcal F}$ must be
at least quadratic in $\bfdel\rho$, which must vanish on the surface of the 
cell for the TETF formalism to be valid. Then Eq. (\ref{B.2}) becomes
\beqy\label{B.4}
-S = \int\left(\frac{\partial {\mathcal F}}{\partial T}\right)_{\rho}
d^3{\bf r} + \int\left(\frac{\partial \rho}{\partial T}\right)_{\bar{\rho}}
\left\{\left(\frac{\partial {\mathcal F}}{\partial \rho}\right)_T - 
\bfdel. \left(\frac{\partial {\mathcal F}}{\partial \bfdel\rho}\right)_T
\right\}d^3{\bf r}   \quad .
\eeqy
Now if equilibrium holds at each temperature $F$ must be a minimum with respect
to variations in $\rho({\bf r})$, which must therefore satisfy the 
Euler-Lagrange equation,
\beqy\label{B.5}
\left(\frac{\partial {\mathcal F}}{\partial \rho}\right)_T -
\bfdel. \left(\frac{\partial {\mathcal F}}{\partial \bfdel\rho}\right)_T
= \lambda \quad ,
\eeqy
where $\lambda$ is a Lagrange multiplier. Thus Eq. (\ref{B.4}) becomes
\beqy\label{B.6}
-S = \int\left(\frac{\partial {\mathcal F}}{\partial T}\right)_{\rho}
d^3{\bf r} + \lambda\int
\left(\frac{\partial \rho}{\partial T}\right)_{\bar{\rho}}d^3{\bf r}
= \int\left(\frac{\partial {\mathcal F}}{\partial T}\right)_{\rho}
d^3{\bf r} + \lambda\left(\frac{\partial }{\partial T}\right)_{\bar{\rho}}
\int\rho\,d^3{\bf r}   \quad .
\eeqy
But the second integral here is just the total number of nucleons in the cell 
(for simplicity we consider just one type of nucleon here), and since this is
temperature independent Eq. (\ref{B.6}) reduces to
\beqy\label{B.7}
S = -\int\left(\frac{\partial {\mathcal F}}{\partial T}\right)_{\rho}
d^3{\bf r} \quad .
\eeqy
Eq. (\ref{3.15}) follows at once.

Likewise, for the specific heat we have from Eqns. (\ref{1.3}) and (\ref{3.22})
\beqy\label{B.8}
Ac_v = T\left(\frac{\partial S}{\partial T}\right)_{\bar{\rho}} =
T\int\left(\frac{\partial \sigma}{\partial T}\right)_{\bar{\rho}}d^3{\bf r}
\quad .
\eeqy
Then in exactly the same way as we have derived Eq. (\ref{B.4}), we find
\beqy\label{B.9}
Ac_v = T\int\left(\frac{\partial \sigma}{\partial T}\right)_{\rho}
d^3{\bf r} + T\int\left(\frac{\partial \rho}{\partial T}\right)_{\bar{\rho}}
\left\{\left(\frac{\partial \sigma}{\partial \rho}\right)_T -
\bfdel. \left(\frac{\partial \sigma}{\partial \bfdel\rho}\right)_T
\right\}d^3{\bf r}   \quad .
\eeqy
In the second term here we can write, using Eq. (\ref{3.15}),
\beqy\label{B.10}
\left(\frac{\partial \sigma}{\partial \rho}\right)_T -
\bfdel.\left(\frac{\partial \sigma}{\partial \bfdel\rho}\right)_T
 = \nonumber \\
&-&\left(\frac{\partial }{\partial T}\right)_{\rho}
\left\{\left(\frac{\partial {\mathcal F}}{\partial \rho}\right)_T -
\bfdel. \left(\frac{\partial {\mathcal F}}{\partial \bfdel\rho}\right)_T
\right\} \quad .
\eeqy
But if we are at equilibrium Eq. (\ref{B.5}) will hold, and both sides of 
Eq. (\ref{B.10}) will vanish, whence Eq. (\ref{B.9}) reduces to
\beqy\label{B.11}
Ac_v = T\int\left(\frac{\partial \sigma}{\partial T}\right)_{\rho}d^3{\bf r}
\quad .
\eeqy
Eq. (\ref{3.26}) follows at once.

\section{Strutinsky-integral theorem}
\label{etfsi}
\renewcommand{\theequation}{C.\arabic{equation}}
\setcounter{equation}{0}

To derive Eqns. (\ref{3.16}) and (\ref{3.17}) we begin by noting that the exact
HF energy of any finite nuclear system (nucleus or WS cell) for Skyrme forces 
and a Slater treatment of the Coulomb exchange energy can be written in the 
local form
\beqy\label{d1}
E_{HF} \equiv E_{HF}[\rho, \tau, {\bf J}]  
= \int\mathcal{E}\left\{\rho({\bf r}), \bfdel\rho({\bf r}),
\tau({\bf r}),{\bf J}({\bf r}) \right\} d^3{\bf r} \quad ,
\eeqy  
where in terms of the exact HF s.p. functions $\phi_i^{HF}({\bf r})$ (not to be
confused with the eigensolutions $\phi_i({\bf r})$ of Eq. (\ref{3.18})) we have
\bmlet
\beqy\label{d2a}
\rho({\bf r}) = \sum_in_i|\phi_i^{HF}({\bf r})|^2 \quad ,
\eeqy
\beqy\label{d2b}
\tau({\bf r}) = \sum_in_i|\bfdel\phi_i^{HF}({\bf r})|^2 
\eeqy
and
\beqy\label{d2c}
{\bf J({\bf r})} = -i\sum_in_i\phi_i^{HF*}({\bf r})\bfdel\times
\mbox{\boldmath$ \sigma$}\phi_i^{HF}({\bf r}) \quad ;
\eeqy
\emlet
for simplicity we do not distinguish here between the two charge states.
For the exact HF quantities $\rho({\bf r}), \tau({\bf r})$ and $\bf J({\bf r})$
let us now write
\bmlet
\beqy\label{d3a}
\rho = \tilde{\rho} +\delta\rho \quad ,
\eeqy
\beqy\label{d3b}
\tau = \tilde{\tau} + \delta\tau 
\eeqy
and
\beqy\label{d3c} 
{\bf J} = \tilde{{\bf J}} +\delta{\bf J} \quad ,
\eeqy
\emlet
where $\tilde{\rho}, \tilde{\tau}$ and $\tilde{{\bf J}}$ represent the smooth
quantities emerging from the (T)ETF calculation. Then to first order in 
$\delta\rho, \delta\tau$ and $\delta{\bf J}$ we have
\beqy\label{d4}
E_{HF}&=& \int{\mathcal E\left\{\tilde{\rho}({\bf r}), 
\bfdel\tilde{\rho}({\bf r}), \cdots, \tilde{\tau}({\bf r}),
\tilde{{\bf J}}({\bf r}) \right\}} d^3{\bf r} \nonumber \\
&+& \int\left\{\left(\frac{\delta\mathcal E}{\delta\rho}\right)_{\tilde{\rho}, 
\tilde{\tau},\tilde{{\bf J}}}\delta\rho 
+ \left(\frac{\delta\mathcal E}{\delta\tau}\right)_{\tilde{\rho},
\tilde{\tau},\tilde{{\bf J}}}\delta\tau
+ \left(\frac{\delta\mathcal E}{\delta{\bf J}}\right)_{\tilde{\rho},
\tilde{\tau},\tilde{{\bf J}}}\cdot\delta{\bf J}
\right\}d^3{\bf r}   \quad ,
\eeqy
where for the functional derivatives appearing here we have
\bmlet
\beqy\label{d5a} 
\left(\frac{\delta\mathcal E}{\delta\rho}\right)_{\tilde{\rho},
\tilde{\tau},\tilde{{\bf J}}} = \tilde{U}({\bf r}) \quad ,
\eeqy
\beqy\label{d5b}
\left(\frac{\delta\mathcal E}{\delta\tau}\right)_{\tilde{\rho},
\tilde{\tau},\tilde{{\bf J}}} = \frac{\hbar^2}{2\widetilde{M^*}({\bf r})}   
\eeqy
and
\beqy\label{d5c}
\left(\frac{\delta\mathcal E}{\delta{\bf J}}\right)_{\tilde{\rho},
\tilde{\tau},\tilde{{\bf J}}} = {\bf \tilde{W}}({\bf r}) \quad ,
\eeqy
\emlet
in which $\tilde{U}, \hbar^2/(2\widetilde{M^*})$ and $\tilde{\bf W}$ are the
smoothed ETF fields appearing in Eq. (\ref{3.18}). Thus Eq. (\ref{d4}) becomes
\beqy\label{d6}
E_{HF} = E_{ETF} + \int\left\{\tilde{U}({\bf r}) \delta\rho +
\frac{\hbar^2}{2\widetilde{M^*}({\bf r})}\delta\tau
+ {\bf \tilde{W}}({\bf r})\cdot\delta{\bf J}\right\}d^3{\bf r} \quad .
\eeqy 
Comparing with Eq. (\ref{3.16}), we see that we now
have to identify the integral in this last equation with the shell correction 
$E_{sc}$ of Eq. (\ref{3.17}).

We next replace the exact HF s.p. functions $\phi_i^{HF}({\bf r})$ in
Eqns. (\ref{d2a}), (\ref{d2b}) and (\ref{d2c}) by the eigensolutions 
$\phi_i({\bf r})$  to Eq. (\ref{3.18}), and define thereby the quantities 
$\rho^\prime, \tau^\prime$ and ${\bf J}^\prime$, respectively; these quantities
will certainly be fluctuating. We can then write
\beqy\label{d7}
\delta\rho = (\rho^\prime - \tilde{\rho}) + (\rho - \rho^\prime) \quad ,
\eeqy
and likewise for $\delta\tau$ and $\delta{\bf J}$. We recall now that 
$\tilde{\rho}$ is the initial approximation (ETF) to the exact HF density 
$\rho$, while $\rho^\prime$ represents our attempt to improve on this 
approximation. Thus the first term on the right-hand side of Eq. (\ref{d7}) 
can be regarded as the first-order estimate of the correction to 
$\tilde{\rho}$, while the second term represents the residual error. Accepting,
then, that our method is essentially one of first-order perturbation theory, we
simply drop the second term of Eq. (\ref{d7}). Treating $\delta\tau$ and 
$\delta{\bf J}$ in the same way, Eq. (\ref{d6}) reduces to
\beqy\label{d8}
E_{HF} = E_{ETF} + \int\left\{\tilde{U}({\bf r})(\rho^\prime - \tilde{\rho}) +
\frac{\hbar^2}{2\widetilde{M^*}({\bf r})}(\tau^\prime - \tilde{\tau})
+ {\bf \tilde{W}}({\bf r})\cdot({\bf J}^\prime - \tilde{{\bf J}})
\right\}d^3{\bf r} \quad .
\eeqy 
But it follows from Eq. (\ref{3.18}) that
\beqy\label{d9}
\int\left\{\tilde{U}({\bf r})\rho^\prime +
\frac{\hbar^2}{2\widetilde{M^*}({\bf r})}\tau^\prime 
+ {\bf \tilde{W}}({\bf r})\cdot{\bf J}^\prime \right\}d^3{\bf r} 
= \sum_in_i\widetilde{\epsilon_i} \quad  .
\eeqy 
Then Eq. (\ref{d8}) reduces to 
\beqy\label{d10}
E_{HF} = E_{ETF} + \sum_in_i\widetilde{\epsilon_i}
- \int\left\{\tilde{U}({\bf r})\tilde{\rho} +
\frac{\hbar^2}{2\widetilde{M^*}({\bf r})}\tilde{\tau}
+ {\bf \tilde{W}}({\bf r})\cdot\tilde{{\bf J}})
\right\}d^3{\bf r} \quad .
\eeqy 
This completes the proof that to first order the shell correction is given
by Eq. (\ref{3.17}).
 
\section{Strong-degeneracy limit of the TETF expansion for entropy}
\label{degen}
\renewcommand{\theequation}{D.\arabic{equation}}
\setcounter{equation}{0}

In the limit of low temperature and high density the TETF expansion of the
entropy density $\sigma_q$ is as follows. For the Thomas-Fermi approximation
we have 
\bmlet
\beqy\label{deg1a}
\sigma_q^{(TF)} = \pi^2\frac{M_q}{\hbar^2}\frac{1}{(3\pi^2\rho_q)^{2/3}}
\frac{\rho_q}{f_q}\,T + O(T^3) \quad .
\eeqy
The first-order correction in $\hbar^2$ to this is
\beqy\label{deg1b}
\sigma_q^{(2)} &=& -\frac{\pi^2}{27}\frac{M_q}{\hbar^2}
\frac{1}{(3\pi^2\rho_q)^{4/3}} 
\frac{1}{f_q}\left\{ \frac{(\mbox{\boldmath$\nabla$}\rho_q)^2}{\rho_q} +
\frac{9}{4}\rho_q\left(\frac{\mbox{\boldmath$\nabla$}f_q}{f_q}\right)^2
+\frac{3}{f_q}\mbox{\boldmath$\nabla$}\rho_q\cdot\mbox{\boldmath$\nabla$}f_q 
\right\}\,T \nonumber \\ 
&+& O(T^3) \quad ,
\eeqy
while the second--order correction in $\hbar^2$ is
\beqy\label{deg1c}
\sigma_q^{(4)} = -\frac{\pi^2}{1620}\frac{M_q}{\hbar^2}\frac{1}
{(3\pi^2\rho_q)^{2}}
\left(17G_1^q - \frac{413}{12}G_2^q +\frac{47}{3}G_3^q \right)\,T 
+ O(T^3) \quad  .
\eeqy
\emlet
In these equations we have defined $f_q =  M_q/M_q^*$, while the $G_i^q$ are
defined in Eqns. (\ref{cv.2a}), (\ref{cv.2b}) and (\ref{cv.2c}).

\newpage
\begin{table}
\centering
\caption{Convergence of the TETF expansion at $\bar{\rho}$ = 0.06 
nucleons.fm$^{-3}$ for cell with $Z = 30, A = 1123$. Temperature $T$ in MeV.}
\label{tab1}
\vspace{.5cm}
\begin{tabular}{|c|cc|cc|cc|}
\hline
$T$ & 0.1 &&  1.0 && 3.0& \\
\hline
        & 0.1943E+02 &&0.1934E+02&&0.1845E+02&\\
$\phi_n$& -0.1027E-01 &19.42&-0.9452E-02&19.33&-0.3529E-02&18.45\\
        & 0.3566E-03 &&0.2959E-03&&0.8834E-04&\\
\hline
        & 0.1818 &&0.1659E+00&&-0.2328E-01&\\
$\phi_p$& 0.3598E-02 &0.1877&0.5124E-02&0.1724&0.2516E-02&-0.02054\\
        & 0.2299E-02 &&0.1404E-02&&0.2167E-03&\\
\hline
        & 0.1477E-01 &&0.1476E+00&&0.4401E+00 &\\
$s_n$ &-0.1387E-05 &0.01477&-0.1318E-04&0.1476&-0.1879E-04&0.4401\\
        & -0.3124E-07 &&-0.2757E-06&&-0.3389E-06&\\
\hline
        & 0.2110E-02 &&0.1695E-01&&0.6378E-01&\\
$s_p$& -0.1537E-02 &-1.005E-03&-0.2133E-02&0.01516&-0.4283E-03&0.06337\\
        & -0.1578E-02 &&0.3437E-03&&0.2441E-04&\\
\hline
$s_p^{s.p.}$&&0.0036233&& 0.01646&&0.06446\\
\hline
        & 0.1477E-01 &&0.1473E+00&&0.4316E+00&\\
$c_{v,n}$&-0.1388E-05 &0.01477& -0.1352E-04&0.1473& -0.2503E-04&0.4316\\
        &-0.3125E-07 && -0.2880E-06&&-0.5186E-06&\\
\hline
        & 0.1895E-02 &&0.1372E-01&&0.3366E-01 &\\
$c_{v,p}$&-0.9657E-03  &0.0075&0.4703E-03&0.01379& 0.4400E-03&0.03410\\
        & 0.6571E-02 &&-0.4022E-03&&-0.2655E-05&\\
\hline
\end{tabular}
\end{table}

\newpage
\begin{table}
\centering
\caption{Convergence of the TETF expansion at $\bar{\rho} = 3 \times 10^{-4}$
nucleons.fm$^{-3}$ for cell with $Z = 40, A = 150$. Temperature $T$ in MeV.}
\label{tab2}
\vspace{.5cm}
\begin{tabular}{|c|cc|cc|cc|}
\hline
$T$ & 0.1 &&  1.0 && 3.0& \\
\hline
        & 0.1472E+02 &&0.1164E+02 &&-0.2114E+01&\\
$\phi_n$& -0.2021E+00  &14.57& -0.1824E+00&11.52&-0.5707E-01&-2.135\\
        & 0.4855E-01 &&0.5564E-01&&0.3647E-01&\\
\hline
        & 0.4465E+01 &&0.4699E+01&&0.6777E+00 &\\
$\phi_p$& -0.3411E-01 &4.496&-0.3642E-01&4.727& -0.1831E-01&0.6960\\
        & 0.6509E-01 && 0.6379E-01&&0.3655E-01&\\
\hline
        & 0.2082E+00  &&0.1026E+01&&0.2923E+01 &\\
$s_n$ &-0.8479E-02 &0.1936& -0.2113E-01&1.006& -0.1382E-01&2.909\\
        &-0.6098E-02 &&0.1083E-02&&0.1035E-03&\\
\hline
        & 0.7193E-02  && 0.6150E-01&&0.8998E+00&\\
$s_p$&  -0.1315E-01 &0.01024&-0.2234E-01&0.0405&-0.1353E-01&0.8864\\
        &0.1620E-01 &&0.1303E-02&&0.1074E-03&\\
\hline
$s_p^{s.p.}$&&0.1199E-03&& 0.06322&&0.7088\\
\hline
        & 0.1385E+00 &&0.3958E+00&& 0.8998E+00&\\
$c_{v,n}$&-0.6991E-02 &0.1410& -0.4233E-02 &0.3905&0.1726E-02&0.9015\\
        &0.9451E-02 && -0.1176E-02&&-0.1625E-03&\\
\hline
        & 0.6965E-02 &&0.5660E-01&&0.2593E+00 &\\
$c_{v,p}$&-0.4111E-02   &-0.00867&-0.2989E-02 & 0.05094&0.2649E-02&0.2619\\
        & -0.1152E-01 &&-0.2666E-02&&-0.3149E-04&\\
\hline
\end{tabular}
\end{table}

\newpage
\begin{table}
\centering
\caption{TETFSI results for number of protons $Z$ and total number of nucleons
$A$ in WS cell for nuclear and beta equilibrium at $T = 0$ as a function of 
$\bar{\rho}$ for force BSK14. TETF results in parentheses. Last two columns
show TETFSI values of pressure $P$ and 
$\Big(\frac{\partial P}{\partial T }\Big)_{\bar{\rho}}$}
\label{tab3}
\vspace{.5cm}
\begin{tabular}{|c|cc|c|c|}
\hline
$\bar{\rho}$ (fm$^{-3}$) & Z& A&$P$ (MeV.fm$^{-3}$)&
$\Big(\frac{\partial P}{\partial T }\Big)_{\bar{\rho}}$ (fm$^{-3}$)  \\
\hline
0.0003 &50 (38)& 200 (146)&0.000940&0\\
0.001 &50 (39) &460 (385)&0.00179&0\\
0.005 &50 (39) &1140 (831)&0.00813&0\\
0.01 & 40 (38) &1215 (1115)&0.0185&0 \\
0.02 & 40 (35) & 1485 (1302)&0.0448&0\\
0.03 & 40 (33) & 1590 (1303)&0.0784&0\\
0.04 & 40 (31) & 1610 (1261)&0.121&0 \\
0.05 & 20 (30) & 800 (1171)&0.175&0 \\
0.06 & 20 (29) & 780(1105)&0.243&0 \\
\hline
\end{tabular}
\end{table}

\begin{table}
\centering
\caption{As for Table \ref{tab3} but with $T$ = 0.1 MeV.}
\label{tab4}
\vspace{.5cm}
\begin{tabular}{|c|cc|c|c|}
\hline
$\bar{\rho}$ (fm$^{-3}$) & Z& A&$P$ (MeV.fm$^{-3}$)&
$\Big(\frac{\partial P}{\partial T }\Big)_{\bar{\rho}}$ (fm$^{-3}$)  \\
\hline
0.0003 &50 (38)& 200 (147)&0.000946&0.0000423\\
0.001 &50 (39) &460 (341)&0.00179&0.000140\\
0.005 &50 (38) &1130 (842)&0.00816&0.000270\\
0.01 & 40 (38) &1210 (1107)&0.0185&0.000346 \\
0.02 & 40 (35) & 1480 (1294)&0.0448&0.000444\\
0.03 & 40 (33) & 1595 (1303)&0.0784&0.000511\\
0.04 & 40 (31) & 1610 (1242)&0.121&0.000568 \\
0.05 & 20 (30) & 800 (1190)&0.175&0.000617 \\
0.06 & 20 (29) & 765 (1116)&0.243&0.000662 \\
\hline
\end{tabular}
\end{table}

\begin{table}
\centering
\caption{As for Table \ref{tab3} but with $T$ = 1.0 MeV.}
\label{tab5}
\vspace{.5cm}
\begin{tabular}{|c|cc|c|c|}
\hline
$\bar{\rho}$ (fm$^{-3}$) & Z& A&$P$ (MeV.fm$^{-3}$)&
$\Big(\frac{\partial P}{\partial T }\Big)_{\bar{\rho}}$ (fm$^{-3}$)  \\
\hline
0.0003 &46 (37)& 310 (234)&0.000633&0.000182\\
0.001 &46 (38) &520 (450)&0.00192&0.000631\\
0.005 &44 (39) &1020 (858)&0.00936&0.00233\\
0.01 & 42 (37) &1280 (1120)&0.0202&0.00329 \\
0.02 & 40 (36) & 1480 (1307)&0.04701&0.00434\\
0.03 & 38 (33) & 1505 (1301)&0.0810&0.00502\\
0.04 & 36 (31) & 1450 (1232)&0.124&0.00553 \\
0.05 & 34 (30) & 1340 (1165)&0.179&0.00568\\
0.06 & 26 (29) & 985 (1082)&0.246&0.00445\\
\hline
\end{tabular}
\end{table}

\vspace*{10.0cm}

\newpage

\end{document}